\documentclass[useAMS,usenatbib]{openjournal}

\usepackage{times,graphicx,amssymb,natbib}
\pdfoutput=1

\newcommand{\msol}{M_{\rm \odot}}
\newcommand{\msolyr}{\rm{\msol \, yr^{-1}}}

\newcommand{\mjup}{M_{\rm Jup}}
\newcommand{\mjupyr}{\rm{\mjup \, yr^{-1}}}

\shorttitle{Self-gravitating circumfragmentary discs}
\shortauthors{D. Forgan}

\begin{document}

\title{Daughter fragmentation is unlikely to occur in self-gravitating circumstellar discs}

\author{Duncan Forgan\altaffilmark{1}}
\affil{Scottish Universities Physics Alliance (SUPA), School of Physics and Astronomy, University of St Andrews, North Haugh, KY16 9SS, UK}
\email{dhf3@st-andrews.ac.uk}

\label{firstpage}

\begin{abstract}

\noindent Circumstellar discs are thought to be self-gravitating at very early times.  If the disc is relatively cool, extended and accreting sufficiently rapidly, it can fragment into bound objects of order a few Jupiter masses and upwards.  Given that the fragment's initial angular momentum is non-zero, and it will continue to accrete angular momentum from the surrounding circumstellar disc, we should expect that the fragment will also possess a relatively massive disc at early times.  Therefore, we can ask: \emph{is disc fragmentation a hierarchical process? Or, can a disc fragment go on to produce its own self-gravitating circumfragmentary disc that produces daughter fragments?}

We investigate this using a set of nested 1D self-gravitating disc models.  We calculate the radial structure of a marginally stable, self-gravitating circumstellar disc, and compute its propensity to fragmentation.  We use this data to construct the local fragment properties at this radius.  For each circumstellar disc model that results in fragmentation, we then compute a marginally stable self-gravitating circumfragmentary disc model.

In general, the circumfragmentary discs are geometrically thick and truncated inside the Hill radius, and are hence stable against daughter fragmentation.  The typical steady-state accretion rate is between 0.1 and 10 percent of the local circumstellar disc accretion rate.  The lifetime of the circumfragmentary discs' self-gravitating phase is somewhat less than 0.1 Myr, quite comparable with that of the circumstellar disc.  We should therefore expect that disc fragments will not produce satellites via gravitational instability, but equally the self-gravitating phase of circumfragmentary discs is likely to affect the properties of satellites subsequently formed via core accretion.


\end{abstract}

\section{Introduction}

\noindent Circumstellar discs are ubiquitous in star formation.  This is a consequence of excess angular momentum in the molecular gas that undergoes gravitational collapse to form a protostellar system.  During the earliest epochs of the star-disc system, it is expected that the mass of the disc is comparable to the mass of the star.  In fact, in the ``first-core/second-core'' paradigm of low mass star formation \citep{Larson1969,Masunaga_1} the first core is typically the progenitor of the disc, with the second-core forming the protostar as $H_2$ dissociation proceeds (e.g. \citealt{Bate2010}).

We must therefore consider disc self-gravity when attempting to understand the evolution of these very young protostellar systems \citep{Lin1987,Laughlin1994}.  Massive discs are susceptible to gravitational instabilities if the Toomre parameter \citep{Toomre_1964}:

\begin{equation} 
Q = \frac{c_s \kappa}{\pi G \Sigma} \sim Q_{crit} \approx 1, 
\end{equation}

\noindent where $c_s$ is the local sound speed, $\Sigma$ is the surface density and $\kappa$ is the epicyclic frequency (for Keplerian discs, this is equal to the angular frequency $\Omega$).  This result applies primarily to axisymmetric perturbations in thin discs, and is usually extended to non-axisymmetric perturbations by changing $Q_{crit}$ to $1.5-1.7$, a result established empirically by numerical simulations (see \citealt{Durisen_review} for a review).  If the disc is geometrically thick, the dispersion relation that governs Toomre instability is modified by a factor $1/(1+ kH)$, where $H$ is the disc scale height and $k$ is the radial wavenumber.  In essence, the disc thickness dilutes the gravitational potential, and hence $Q_{crit}$ decreases below 1 (see e.g. \citealt{Kim2007}).

If the above criterion is satisfied, gravitational instabilities in the disc produce spiral density waves.  This action heats the disc through stress heating and low Mach number shocks, which increase the local sound speed and hence $Q$.  The competition between this heating and radiative cooling (which reduces $Q$) produces self-regulation of the instability, resulting in marginally stable self-gravitating discs \citep{Paczynski1978}, where the value of $Q$ is maintained near to $Q_{crit}$.  

The continuous production of non-axisymmetric structures in the marginally stable regime produces gravitoturbulence \citep{Gammie}.  This gravitoturbulence can be parametrised using pseudo-viscous prescriptions, as is commonly done for other turbulent processes in discs such as the magneto-rotational instability (MRI) (\citealt{Blaes1994,Balbus1999}, see also combined gravo-magneto disc instability calculations such as \citealt{Martin2014}, which are relevant for discs at the end of the self-gravitating phase).  Typically, pseudo-viscous approaches use the dimensionless $\alpha$-parametrisation, which allows modellers to describe the viscosity using the disc's local velocity and length scales \citep{Shakura_Sunyaev_73}.  This has been an extremely useful tool for building self-consistent steady state self-gravitating disc models \citep{Rafikov_05,Clarke_09,Rice_and_Armitage_09} as well as evolutionary models of the star-disc-envelope system \citep{Lynden-Bell1974,Pringle1981a,Lin1990,Rice2010}.  These models are acceptable descriptions of the angular momentum transport in the disc provided that the spiral density waves excited by the instability do not exert long range gravitational forces on the disc, i.e. the gravitational stress on the disc is locally determined \citep{Lodato_and_Rice_04,Forgan2011}.  This is typically the case if the disc to star mass ratio is low, and the disc scale height is relatively small.

In the pseudo-viscous approximation, we can find regimes where the regulatory mechanisms that produce marginally stable self-gravitating discs fail, and the disc fragments into bound objects.  \citet{Gammie} showed that if the disc's radiative cooling is sufficiently strong, the gravitational stress saturates, breaking self-regulation and producing fragments.  This criterion is related to the local cooling time, $t_{cool}$, and is usually expressed using the dimensionless cooling time parameter $\beta_{c}=t_{cool} \Omega$, i.e.

\begin{equation} \beta_{c} \leq \beta_{crit}, \end{equation}

\noindent where the exact value of $\beta_{crit}$ is typically established by numerical simulations.  Using the 2D shearing sheet approximation, \citet{Gammie} showed that when the ratio of specific heats in two dimensions $\gamma_{2D}=2$, $\beta_{crit} = 3$.  \citet{Rice_et_al_05} extended this analysis to 3D smoothed particle hydrodynamics (SPH) simulations (where $\beta_{c}$ was fixed at some value), and explored the dependence of $\beta_{crit}$ on $\gamma$.  They were able to confirm Gammie's assertion that, in the case where the disc has reached thermal equilibrium (i.e. the radiative cooling and stress heating due to gravitational instability are in balance), the fragmentation criterion can be recast in terms of the $\alpha$ parameter:

\begin{equation} 
\alpha = \frac{4}{9 \gamma(\gamma-1)\beta_c} > \alpha_{crit}
\end{equation} 

\noindent where  $\alpha_{crit}\approx 0.06$ for all values of $\gamma$.  The fragmentation criterion can now be interpreted as a maximum, saturating GI stress that the disc can support in thermal equilibrium without fragmentation.  

The precise value of this saturating $\alpha$ has been called into question by studies of the convergence of 3D SPH simulations \citep{Meru2011}.  Using standard SPH numerical experiments where $\beta_c$ is held fixed, the measured $\beta_{crit}$ increases (i.e. $\alpha_{crit}$ decreases) as the total number of particles is increased.  This is in spite of the fact that the local Jeans mass is already well resolved \citep{Forgan2011a}.  The community has proposed several explanations for this non-convergence, including poor resolution of the disc's vertical structure \citep{Lodato2011} and the implementation of fixed $\beta$-cooling \citep{Rice2012,Rice2014}.  The fact that grid-based simulations do not show increasing $\beta_{crit}$ with increased resolution, and a saturating $\alpha$ of order 0.1, suggests that this is indeed a numerical problem associated with SPH \citep{Michael2012}.

The inner regions of discs are typically too hot for $Q\sim 1$, and are therefore not gravitationally unstable and incapable of fragmentation.  Also, if the primary contribution to $\alpha$ is from GI stress, it will typically decrease with increasing proximity to the central star \citep{Armitage_et_al_01,Rice_and_Armitage_09,Rice2010}.  These facts immediately suggest a minimum radius at which fragmentation can occur.  Semi-analytic methods, grid-based and particle based simulations, tend to agree that for self-gravitating protostellar discs the minimum is typically greater than $\sim 40$ au, using typical disc parameters \citep{Rafikov_05, Matzner_Levin_05, Whit_Stam_06,Mejia_3,Stamatellos2008, intro_hybrid, Clarke_09}.  

However, most studies of disc fragmentation do not consider the consequences of the fragment's angular momentum.  Disc fragments will contain some relic angular momentum from their formation, while continuing to accrete angular momentum from its local disc environment.  High-resolution simulations of the fragment post-formation suggest that it rotates as a solid body with an angular velocity around 10\% of its break-up speed \citep{Nayakshin2011a}.  It is likely that a circumfragmentary disc will form (see e.g. \citealt{Boley2010b,Galvagni2012}) and regulate the fragment's accretion in a microcosm of the star-circumstellar disc relationship.  

Again, if we consider disc fragment collapse to be broadly similar to protostellar collapse, we may expect the circumfragmentary disc to possess a mass of order that of its host.  This gives rise to the possibility that the circumfragmentary disc could itself fragment, producing satellites for the host fragment.  We will henceforth refer to this as \emph{daughter fragmentation}.

Daughter fragmentation has clear consequences for the properties of subsequent satellites, as well as the initial conditions for satellite formation via core accretion \citep{Canup2002,Canup2006,Mosqueira2003,Mosqueira2003a}.  It has been shown that the interior of disc fragments can act as grain processing factories \citep{Nayakshin2012}, and this processed material can be bequeathed to the local environment if a fragment is tidally disrupted.  We might therefore expect that if a planet/brown dwarf is formed via disc fragmentation, then its satellites will have distinctive properties compared to a satellite formed around a core-accretion planet, either because a) the satellite is itself a daughter fragment, or b) is assembled from the remains of a daughter fragment, which has processed the disc material in its interior.  But is daughter fragmentation likely?

Hydrodynamic simulations of circumfragmentary discs \citep{Shabram2013} have shown that this initially large disc mass is quickly accreted by the host, in a manner quite analogous to massive protostellar discs \citep{Forgan2011}.  The circumfragmentary disc is typically geometrically thick, with aspect ratio $H/r > 0.2$ and as large as 0.75!  The discs are truncated at approximately one third of the Hill Radius:

\begin{equation}
R_H = a \left(\frac{M_{frag}}{3M_*}\right)^{1/3}
\end{equation}

\noindent Where $a$ is the fragment's orbital semimajor axis, $M_{frag}$ is the mass of the fragment including its disc, and $M_*$ is the host star mass.  The discs are stable against fragmentation, despite being able to accrete from the circumstellar disc at relatively high rates for a brief interval.

Semi-analytic models of MRI-active circumplanetary discs with an outer self-gravitating component \citep{Keith2014} also indicate that fragmentation is unlikely, with large aspect ratios and typical values of $\beta_c$ around $10^5$.  However, this work did not make assumptions regarding the formation mechanism of the planet-disc system, and it is possible that as for protostellar discs, circumfragmentary discs are very weakly ionised at early times and hence not magnetically active.  Most other studies of circumplanetary discs do not consider the mass regimes we are interested in, and focus specifically on lower mass planet-disc systems (e.g. \citealt{Quillen1998,DAngelo2002, Ayliffe2009,Ward2010,Martin2011}).

While the evidence suggests daughter fragmentation is unlikely, the parameter space that this process could inhabit is yet to be fully explored.  We should hence consider what conditions are required for disc fragments, with masses near the planet/brown dwarf boundary, to produce a fully self-gravitating circumfragmentary disc, and whether there is any part of the parameter space for such a disc that results in the production of daughter fragments.  

This work investigates the phenomenon of circumfragmentary discs using a set of nested self-consistent 1D self-gravitating disc models.  We compute self-gravitating circumstellar disc models, which deliver the initial conditions for disc fragmentation.  We then compute self-gravitating circumfragmentary disc models based on these constraints, and study their properties, including their propensity to produce daughter fragments.

The nested disc models are described in section \ref{sec:methods}; we display and discuss results from these models in section \ref{sec:results}, and summarise the work in section \ref{sec:conclusions}.

\section{Method} \label{sec:methods}
\subsection{Self-gravitating circumstellar disc models}

\noindent We construct self-consistent self-gravitating circumstellar disc (CSD) models as described in \citet{Forgan2011a}.  We fix the disc's accretion rate,

\begin{equation}
 \dot{M}_{CSD} =\frac{3 \pi \alpha c^2_{s} \Sigma}{\Omega}
\end{equation} 

\noindent as constant throughout the disc extent, and we demand a fixed 
 
\begin{equation} 
 Q = \frac{c_s \Omega}{\pi G \Sigma} = 1.5.
 \end{equation}
 
\noindent We calculate $\Omega$ assuming the disc is Keplerian, i.e:

\begin{equation}
\Omega(r) = \sqrt{\frac{G M_{enc}(r)}{r^3}}
\end{equation}

\noindent Where $M_{enc}(r)$ is the total enclosed mass (star + disc) within $r$.  The local pseudo-viscous $\alpha$ parameter is calculated assuming local thermodynamic equilibrium:
 
\begin{equation}
 \alpha = \frac{4}{9\gamma(\gamma-1) \beta_{c}}
\end{equation}

\noindent The local cooling time (and hence $\beta_c$) can be determined given the cooling function:

\begin{equation}
\dot{u} = \frac{\sigma_{SB} T^4}{\tau + 1/\tau}
\end{equation}

\noindent Where the denominator is constructed to allow smooth transition between optically thick and optically thin regimes \citep{Hubeny1990}.  The optical depth $\tau = \Sigma \kappa$, where $\kappa$ is the opacity.  Assuming

\begin{equation}
\rho = \frac{\Sigma}{2H}
\end{equation}

\noindent and an appropriate equation of state to define $\kappa$, the system can be closed, and appropriate iteration on values of $\Sigma$ produces solutions for the disc at any radius.  We assume that $\alpha$ cannot exceed a saturation value $\alpha_{sat}$, which we set to a cautious value of 0.1.  We employ the same equation of state as \citet{Forgan2011a} (which in turn is taken from \citealt{Stam_2007}), where we use the opacity law of \citet{Bell_and_Lin}.

We calculate fragmentation using the Jeans mass criteria of \citet{Forgan2011a}, which also yields the initial fragment mass:

\begin{equation}
M_{frag} = \frac{4\sqrt{2Q} \pi^3}{3G}\frac{c^4_s}{\Sigma\left(1+1/\sqrt{\beta_c}\right)}.
\end{equation}

\noindent This expression relies on the relationship between the local cooling time and surface density perturbations, as was elucidated by \citet{Cossins2008}. $M_{frag}$ is at minimum a few Jupiter masses, and at maximum extends well into the brown dwarf regime (see \citealt{Forgan2011a,Forgan2013} for more).

\subsection{Self-gravitating circumfragmentary disc models}

\noindent The above calculations yield a set of self-gravitating circumstellar disc models, uniquely specified by the star mass $M_*$, the circumstellar accretion rate $\dot{M}_{CSD}$, and the inner and outer disc radii.  For any given $\dot{M}_{CSD}$ and disc radius, we can calculate $M_{frag}$ (if fragmentation occurs), and a corresponding Hill radius $R_H$.  

Given some extra assumptions, this is sufficient data to construct a self-gravitating circumfragmentary disc, where we now replace the central star mass with a central planetary mass.  Therefore, for every combination of self-gravitating circumstellar disc model parameters that results in fragmentation, we can essentially repeat the disc calculation process, under several extra constraints:

\begin{enumerate}
\item The total mass of the fragment-disc system cannot exceed $M_{frag}$,
\item The circumfragmentary disc outer radius must be equal to $R_H/3$
\end{enumerate}

\noindent We cannot \emph{a priori} specify the circumfragmentary disc's initial accretion rate, so we cannot follow the same precise steps as before.

We instead propose a trial accretion rate for the circumfragmentary disc, $\dot{M}_{CFD}$, and then compute the disc corresponding to this accretion rate, extending to $1/3 R_H$.   The discs are generally much thicker and we hence impose a fixed $Q=0.7$ \citep{Kim2007}. If the first constraint is not met, $\dot{M}_{CFD}$ is adjusted and the procedure is repeated.  The second constraint in the above list is of course met by construction. 

This extra iterative step ensures that calculating circumfragmentary discs is an order of magnitude more computationally intensive.   We are also ignorant to the initial mass ratio of the disc to its host, and we must also fix this to satisfy the first constraint.  Following \citet{Shabram2013}, we assume that initially the mass of the CFD is equal to the mass of the central object, i.e. given the total mass of the system, $M_{frag}$:

\begin{equation}
M_{CFD} = 0.5 M_{frag}.
\end{equation}

\noindent Once the CFD model is computed, we test for daughter fragmentation, again using the local Jeans mass criterion as before.

\section{Results \& Discussion} \label{sec:results}

\begin{figure}
\begin{center}
\includegraphics[scale=0.5]{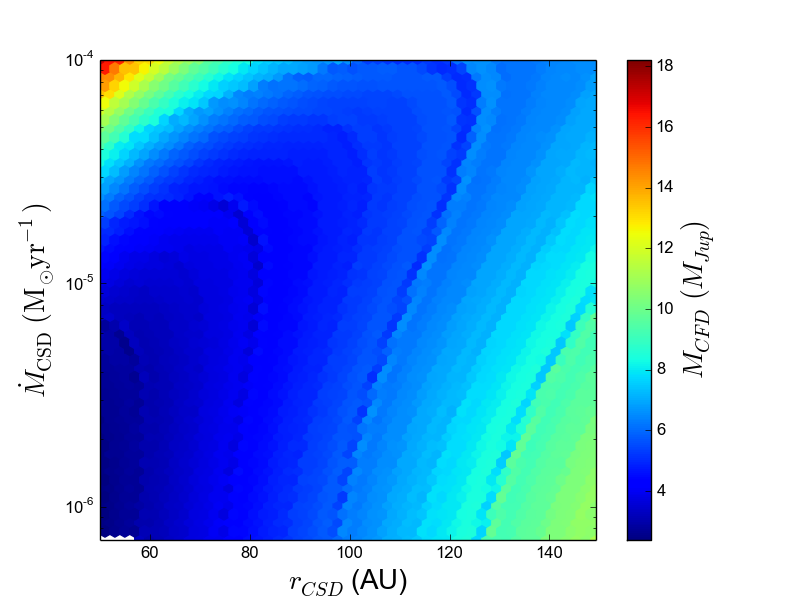} 
\caption{The mass of the circumfragmentary disc, calculated for fragments forming in circumstellar discs with a given accretion rate (y axis), at a given radius (x axis).  This traces the total fragment mass, as we have fixed the CFD mass to be precisely half the total fragment mass.\label{fig:mass_q_05}}
\end{center}
\end{figure}

\noindent We computed a grid of circumstellar disc models for a star of mass $M_*=1\msol$, with constant accretion rates from $10^{-7}-10^{-4} \msolyr$, and disc outer radii from 50 to 150 au.  Figure \ref{fig:mass_q_05} shows the mass of the circumfragmentary disc that results from fragments forming inside circumstellar discs at a given CSD accretion rate and CSD radius.  As we have fixed this property to be precisely half the fragment mass, the resulting plot is very similar to the fragment mass contour plots shown in \citet{Forgan2011a}.  Note that we discard CSD models where fragmentation does not occur, and that we require quite large CSD accretion rates to achieve fragmentation, at least $\sim 10^{-6} \msolyr$.  This underscores the need for fragmentation to occur promptly in the self-gravitating phase of the circumstellar disc.  Such high accretion rates result in significant and rapid disc mass depletion, precluding delayed fragmentation \citep{Forgan2011}.

\begin{figure}
\begin{center}
\includegraphics[scale=0.5]{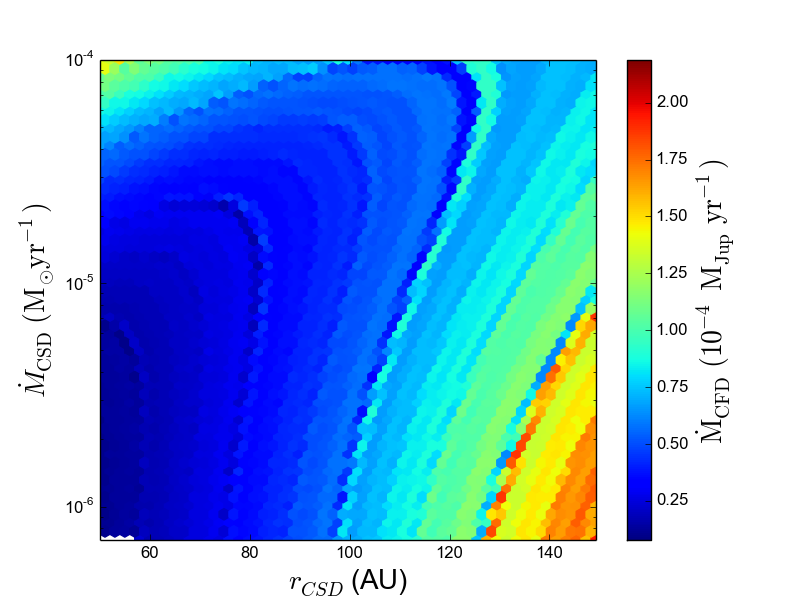} 
\caption{As Figure \ref{fig:mass_q_05}, for the accretion rate of the circumfragmentary disc.\label{fig:mdot_cfd_q_05}}
\end{center}
\end{figure}

This need for relatively large accretion rates to induce fragmentation provides us with a clue as to whether daughter fragmentation is viable.  The accretion rate of the circumfragmentary discs (Figure \ref{fig:mdot_cfd_q_05}) is of order $10^{-4}-10^{-5} \mjupyr$, i.e. $10^{-7}-10^{-8} \msolyr$, which is quite consistent with those measured in 3D hydrodynamic simulation \citep{Shabram2013}, where the true CFD accretion rate is slightly less, and weakly dependent on the rate of infall from the CSD.


\begin{figure}
\begin{center}
\includegraphics[scale=0.5]{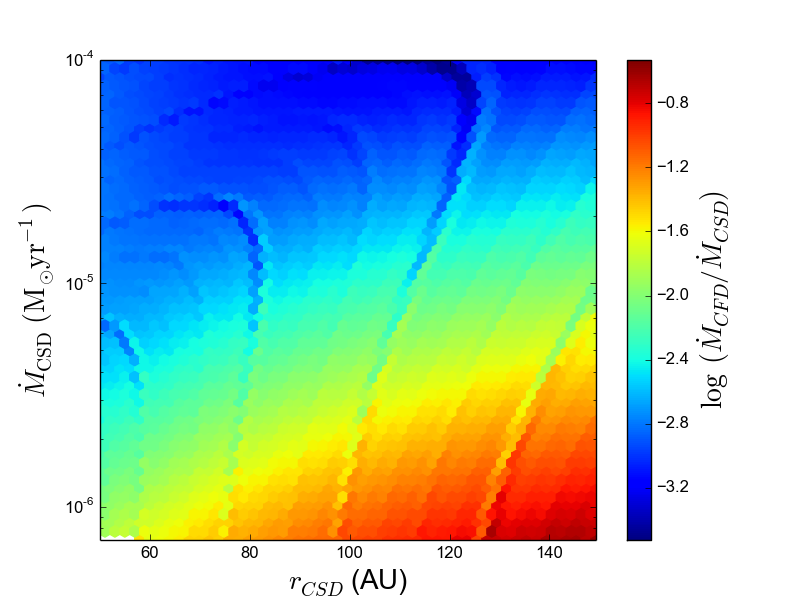} 
\caption{As Figure \ref{fig:mdot_cfd_q_05}, where we now normalise the CFD accretion rate by the local CSD accretion rate.\label{fig:accratio_cfd_q_05}}
\end{center}
\end{figure}

\noindent Comparing the CFD accretion rates with the ambient CSD accretion rate (Figure \ref{fig:accratio_cfd_q_05}) shows that the CFD accretion flow is at least an order of magnitude less than that of the CSD.  However, the CFD's accretion rate is largely controlled by the total fragment mass, which does not vary significantly as $\dot{M}_{CSD}$ is increased, and so the CFD accretion flow can be several orders of magnitude below the CSD flow at high $\dot{M}_{CSD}$.  This mismatch in accretion flow is mainly a symptom of our enforcing a constant $Q$ and constant $M_{frag}$, but in practice we would expect $M_{frag}$ to increase as the fragment accretes from the CSD.  It is important to note that a large fraction of the central object's mass is processed by the CFD, slowing the growth of the fragment and potentially boosting its entropy higher than predicted by``hot-start'' models of protoplanet accretion \citep{Owen2016}.

The fragment's accretion rate will be governed by the fragment's radial migration through the CSD (which we do not model), and by the CSD's own evolution towards an exit from the self-gravitating phase.  As we do not explicitly model any interaction between the CFD and the surrounding CSD environment, we are not equipped to answer this question, but it does seem to be the case that if CFDs were to fragment via substantial infall from the nearby CSD, then this would have to happen rapidly if at all, as both the CFD and CSD rapidly deplete their mass through efficient angular momentum transport.


\noindent We find that the typical disc lifetime:

\begin{equation}
\tau_{CFD}  = \frac{M_{CFD}}{ \dot{M}_{CFD}}
\end{equation}

\noindent is of order 0.2 million years.  The equivalent timescale for the CSD is at most half a million years, so we can see that (assuming no fragment migration, and no growth of the total mass in the CFD system) both the CFD and CSD will leave the self-gravitating phase on similar timescales.  Of course, we must bear in mind that $\tau_{CFD}$ is not the self-gravitating disc lifetime, i.e. the timescale on which the disc is sufficiently massive for self-gravity to dominate the transport.  The self-gravitating phase will end when the mass ratio of disc to central object drops below $\sim 0.1$, which is likely to be significantly less than $\tau_{CFD}$.  

\begin{figure}
\begin{center}
\includegraphics[scale=0.5]{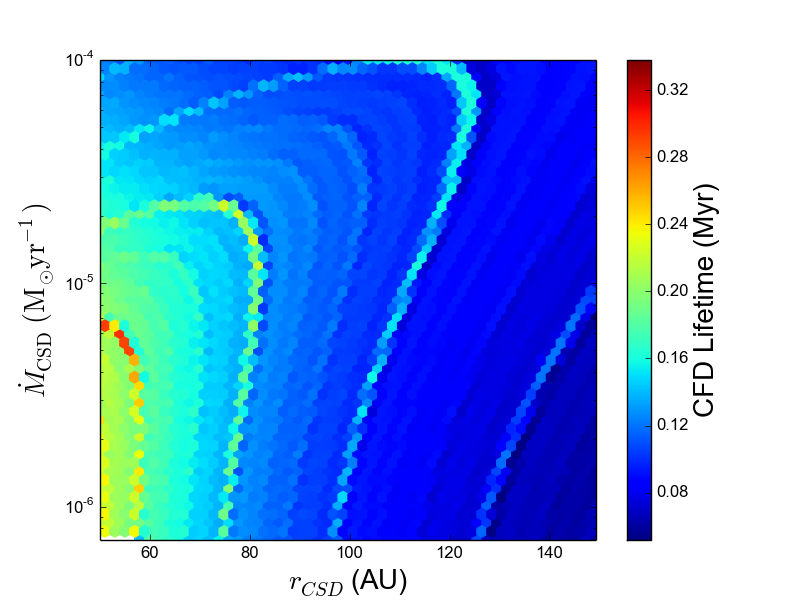} 
\caption{As Figure \ref{fig:mass_q_05}, for the CFD disc lifetime\label{fig:disclife_q_05}}
\end{center}
\end{figure}

Throughout, we find no evidence for daughter fragmentation.  The local Jeans mass criterion is not satisfied for any of the CFD models, mainly due to the disc thickness diluting the gravitational potential.  It is important to note that the disc thicknesses are large enough to potentially invalidate the local transport approximation \citep{Forgan2011}, which is likely to slightly underestimate the true gravitational stress (Hall et al, submitted).  We should therefore expect that simulations of CFDs in this parameter space should show significant non-axisymmetric structures affecting the gravitational potential across the disc's entire extent.  Note that this is completely independent of the non-axisymmetric structures feeding into the CFD beyond the disc truncation radius.

This being the case, how will CFDs affect the future evolution of the fragment into a bound object? Perhaps most crucial in this process is the role of the CFD in the fragment's inward radial migration.  Fragments do appear to migrate inward quite rapidly in self-gravitating CSDs \citep{Baruteau2011}, but most studies of fragment migration do not resolve the fragment's internal structure sufficiently to show a disc.  For low mass planets, the presence of a circumplanetary disc can reduce the migration rate.  This is especially true for thick circumfragmentary discs, as the torques from inner and outer Lindblad resonances decrease with increasing CFD scaleheight \citep{Ward1997, Ayliffe2010}.  It is also clear that the ability of the fragment system to drive a gap will be affected by its CFD, and as a result its migration rate \citep{Malik2015a}.

Even without daughter fragmentation, this brief phase of strong spiral density waves being launched in the circumfragmentary disc (rather than imposed tidally) will have implications for the growth of solid grains.  In self-gravitating CSDs, spiral density waves provide pressure maxima suitable for concentrating metre-sized particles \citep{Rice2004,Gibbons2012}, but eccentricity growth and an increased velocity dispersion may frustrate the growth of planetesimals in such concentrated grain swarms \citep{Walmswell2013}.  In any case, the dynamical landscape for subsequent grain growth is indelibly altered by the spiral density waves launched during the CFD's self-gravitating phase.

We should therefore expect that exomoons are unlikely to be the direct descendants of daughter fragments.  Equally, where brown dwarfs possess planetary mass satellites as opposed to similar mass binary companions (e.g. \citealt{Udalski2015a}), these are also presumably formed via core accretion (cf \citealt{Meru2013b}).  That being said, the self-gravitating disc phase can transform disc chemistry, and those effects are likely to persist into the later planet/moon formation era \citep{Ilee2011,Evans2015}.  We might then expect that if the host body is a disc fragment, then satellites might possess characteristic differences from satellites with hosts formed by core accretion.  Further work in both the radiation hydrodynamics of circumfragmentary discs, and consequent chemical evolution, is required to address this possibility.


\section{Conclusions} \label{sec:conclusions}

\noindent We have conducted a parameter study of fragmentation in self-gravitating circumstellar discs (CSDs), to determine if those fragments are likely to themselves host self-gravitating discs that produce daughter fragments.  These circumfragmentary  discs (CFDs) are qualitatively similar to those produced in high resolution 3D numerical simulations, confirming that our simple $\alpha$-disc models are a suitable approximation for discs where self-gravity is the dominant physical process.

Despite surveying a large region of CSD parameter space, we find no evidence of daughter fragmentation.  The CFDs produced are geometrically thick, and are hence stable against fragmentation.  In a manner similar to CSDs, the CFD accretion rates are relatively large, such that the CFD is likely to leave the self-gravitating phase quite rapidly, and other sources of turbulence will come to dominate the angular momentum transport.  Indeed, we show that the CFD and CSD will exit the self-gravitating phase on 
similar timescales.

Consequently, satellite formation around planets and brown dwarfs by gravitational instability and disc fragmentation appears to be very unlikely.  The only caveat from our analysis is our neglecting of the region just beyond the CFD, where it connects to the CSD.  It may well be the case that the accretion flow of CSD material onto the CFD can be triggered into fragmenting (cf \citealt{Meru2015}).  It seems unlikely such daughter fragments could achieve stable orbits around the parent to become satellites, but this remains an open question, and worthy of further simulation and analysis.

Even if disc fragmentation is not active in the environments surrounding protoplanets and proto-brown dwarfs, the self-gravitating phase of their discs will set the stage for future satellite formation by core accretion. Therefore, their hydrodynamics and chemical evolution require further study if we are to understand the properties of satellites formed around the descendants of disc fragments.

\section{Acknowledgements}

\noindent The author gratefully acknowledges support from the ECOGAL project, grant agreement 291227, funded by the European Research Council under ERC-2011-ADG, PI: Ian Bonnell, and thanks Ken Rice for useful discussions.

\bibliographystyle{mn2e} 
\bibliography{selfgravCPD}

\label{lastpage}

\end{document}